\documentclass[12pt]{article}
\usepackage[pctex32]{graphicx}
\linethickness{0.4pt}

\rm
\def\eps{\epsilon}
\def\var{\varepsilon}
\def\f{\frac}
\def\om{\omega}
\def\S{\Sigma}

\begin{document}
\begin{center}
{\large
\bf Electron self-energy and effective mass  in a single heterostructure \vspace*{2cm}\\}
\hspace*{0cm}
  Xiu-Kun Hua$^{1}$\footnotetext{ electronic mail:
   huaxk@pub.sz.jsinfo.net}, Yin-Zhong Wu,$^{1,2}$, and Zhen-Ya Li$^{3,1}$
\vskip 0.2  true cm
$^{1}${\it Department of Physics, Suzhou University, Suzhou, 215006, China$^*$}\\
$^{2}${\it Department of physics, Changshu College, Changshu, 215500, China}\\
$^{3}${\it CCAST(World Laboratory), P. O. Box 8730, Beijing 100080, China }\\
\vspace*{2cm}
\end{center}
\baselineskip 0.8cm
\begin{center}
{\bf Abstract}\\
\end{center}
\hspace*{0.5cm}
In this paper, we investigate the electron self-energy and effective mass
in a single heterostructure using Green-function method.
 Numerical calculations of the electron self-energy and
effective mass for $GaAs/AlAs$ heterostructure are performed.
The results show that the self energy
(effective mass) of electron, which incorporate the energy of electron coupling to
interface-optical phonons and half three-dimension LO
phonons, monotonically increase(decrease) from that of interface polaron to
that of 3D bulk polaron with the increase of the distance between the position of the electron and interface.
\vspace*{1cm}\\
{\bf Key words:} \hspace*{0.5cm}  Semiconductors, Electron-phonon interactions \\

\newpage 
{\normalsize\bf 1. Introduction}\\
\hspace*{0.5cm}
The heterostructure and superlattice of insulators and semiconductors
have many potential electronic and optical properties, which continue to
receive much attention for both theoretical research and their application on electronic and optical devices[1, 2]. The polaron effects of electron-interface-optical-phonon
interaction result in some changes to the electron self-energy and the effective mass
which are very different from those in 3D system, and have been investigated extensively. In the past years, many works on the properties of interface polaron
 were studied using the perturbation-
theory method [3-5], the Lee-Low-Pines(LLP) variational method [6-8] and the
Feynman path integral approach [9-11].
Hai [5], for example,
investigated the polaron energy and effective mass
in a quantum well by applying the perturbation-theory method,
Li and Gu [7] investigated intermediate-coupling polaron in a
polar-crystal slab by using LLP variational method.
A few works,
which include the study of many-
body effects in the normal-state polaron system [12] and the investigation of
exact ground state properties of polaron in the limit of large dimensions [13],
have been done using the powerful Green-function method. Recently Charrour presented a systematic study of 
the ground state binding energy of  hydrogenic impurity in cylindrical quantum dot [14]
and the effect of  electron-phonon interaction on an electron bound to an impurity in a spherical
quantum dot is studied by Melinikov etc [15].\\
\hspace*{0.5cm}
 In the past, many works have been contributed to investigate the properties of interface polaron[16,17]. However,
rather insufficient attentions have been paid to the properties of polaron as a function of the distance between the position of the electron and interface.
In this paper, we apply the Green-function method to study the effect
of electron-phonon interactions on the self-energy and the effective mass of an electron in a single heterostructure. We put Hamiltonian into the second quantization
 representation only on the area paralleling to interface. Our
numerical results of the self-energy and effective mass using $GaAs/AlAs$
heterostructure as an example show that the electron
self-energy and effective mass in single heterostructure are strongly
related to the distance of electron to interface.
We find that the total electron self-energy,
which incorporate the contribution of electron coupling to
interface-optical phonons and half 3D bulk LO
phonons to electron self-energy, monotonically increase
from that of interface polaron to that of 3D polaron with the
increase of distance of electron to interface.
We also find that the total effective mass,
which incorporate the contribution of electron interacting with
interface-optical phonons and half 3D bulk LO
phonons to the effective mass, monotonically decrease
from that of interface polaron to that of 3D polaron with increase of
distance of electron to interface.\\
{\normalsize\bf  2. Hamiltonian and calculation}\\
\hspace*{0.5cm}
A single heterostructure is taken into account. The space
for $z>0$ is occupied by $GaAs$ (crystal 1) and
for $z<0$ by AlAs (crystal 2).
 Considering only the motion of an electron with mass m,
the Hamiltonian for the coupling of an electron to bulk and interface
optical phonons in a single heterostructure can be given by\\
$$
H=\f{p^2}{2m}+V(z)
+\sum_q\hbar{\om}_{Ln}{a^{\dag}}_qa_q
+\sum_{q_{//}}\hbar{\om}_{\pm}{a^{\dag}}_{q_{//}}a_{q_{//}}
+H_{e-BO}+H_{e-IO},
\eqno(1)
$$
where p is the momentum operator of the electron, m is the band mass of
electron, V(z) is the confining potential in the z direction,
${a_q}^{\dag}(a_q)$ is the creation
(annihilation) operator of a bulk LO phonon with wave vector q and energy
$\hbar{\om}_{Ln}$,
${{a_q}_{//}}^{\dag}({a_q}_{//})$ is the creation
(annihilation) operator of an interface-optical (IO) phonon
with wave vector $q_{//}$ and energy $\hbar{\om}_{\pm}$.
$H_{e-BO}$ is
the electron-LO-phonon interaction Hamiltonian in half 3D bulk crystal,
$H_{e-IO}$ is the electron-IO-phonon interaction Hamiltonian.
We regard the band mass to be homogeneous one.\\
\hspace*{1cm}
The polarizing electron cloud of the interface ions induces an image potential
which is
$$
V(z)=\f{e^2(\eps_{\infty 1}-\eps_{\infty 2})}
       {4z\eps_{\infty n}(\eps_{\infty 1}+\eps_{\infty 2})},
\eqno(2)
$$
where $\eps_{\infty 1}$ and $\eps_{\infty 2}$ are the optical dielectric
constants of crystal 1 and 2, respectively. n=1 and 2 denote crystal 1
and 2.
For a single heterostructure, the electron-LO-phonon interaction Hamiltonian
of an electron with half 3D bulk and interface optical phonons
was given by Mori and Ando
[18]:
$$
H_{e-BO}=\sum_{q_z>0}\sum_{q_{//}}e^{iq_{//}\cdot r_{//}}\Gamma_{Ln}(q_{//},q_z,z)
[a_{q_{//}j}+{a^{\dag}}_{-q_{//}j}]
\eqno(3)
$$
and
$$
H_{e-IO}=\sum_{q_{//}}e^{iq_{//}\cdot r_{//}}\Gamma_{IO}^{\pm}(q_{//},z)
[a_{q_{//}}+{a^{\dag}}_{-q_{//}}],
\eqno(4)
$$
where $\Gamma_{Ln}(q_{//},q_z,z)$ and 
$\Gamma_{IO}^{\pm}(q_{//},z)$
are the coupling functions that describe the coupling
strengths of a single electron with the half 3D bulk optical-phonon
modes in crystal 1 and 2 and with the interface optical phonon
modes at the position z, respectively.
The expression of the coupling function in half 3D bulk crystal
was given by
$$
\Gamma_{Ln}(q_{//},q_z,z)=-\left[\f{\om_{Ln}e^2}{2V}
                           \right]^{1/2}
          \left[\f{1}{\eps_{\infty n}}-\f{1}{\eps_{0n}}\right]^{1/2}
          \f{1}{[{q_{//}}^2+{q_z}^2]^{1/2}}
          \theta_n(z)2sin(q_zz),
\eqno(5)
$$
where $\om_{Ln}$ denote the frequency of LO phonons in polar crystal
n(n=1 and 2), V denotes the volume of polar crystal and
$$
\theta_n(z)=
             \left\{\begin{array}{ll}
             1,\hspace*{1cm}& \mbox{if electron is in crystal n}\\
             0,\hspace*{1cm}& \mbox{other}.
             \end{array}\right.
$$
 $\Gamma_{IO}^{\pm}(q_{//},z)$ was given by
$$
\Gamma_{IO}^{\pm}(q_{//},z)
=-\left[\f{\om_{\pm}e^2}{2S}\right]^{1/2}
 \left[\f{1}{\beta_1^{-1}(\om_{\pm})+\beta_2^{-1}(\om_{\pm})}\right]^{1/2}
 \f{1}{\sqrt{2q_{//}}}e^{-q_{//}|z|},
\eqno(6)
$$
where
$$
\beta_n(\om_{\pm})=\left(\f{1}{\eps_{\infty n}}-\f{1}{\eps_{0 n}}\right)
                   \f{\om_{Ln}^2}{\om_{\pm}^2}
        \left[\f{\om_{\pm}^2-\om_{Tn}^2}{\om_{Ln}^2-\om_{Tn}^2}\right]^2,
        n=1,2
\eqno(7)
$$
and $\om_{\pm}$ is decided by
$$
\left\{\begin{array}{l}
   \eps_n(\om)=\eps_{\infty n}\f{\om^2-\om_{Ln}^2}{\om^2-\om_{Tn}^2}\\
   \eps_1(\om)+\eps_2(\om)=0.
   \end{array}\right.
\eqno(8)
$$
\hspace*{0.5cm}
In order to represent the Hamiltonian in the second quantization representation,
we expand the electron wave function $\Psi$(r) ($\Psi^{\dag}$(r))
in a basis set $\phi_k$(r)
(${\phi_k}^{\dag}$(r)) with the well known method
$$
\Psi(r)=\sum_{k_{//}}c_{k_{//}}\phi_{k_{//}}(r),\\
\eqno(9a)
$$
$$
\Psi^{\dag}(r)=\sum_{k_{//}}{c_{k_{//}}}^{\dag}{\phi_{k_{//}}}^*(r),
\eqno(9b)
$$
where
$$
\begin{array}{lll}
\phi_{k_{//}}(r)&=&\f{1}{2\pi}e^{ik_{//}.r_{//}},\\
{\phi_{k_{//}}}^*(r)&=&\f{1}{2\pi}e^{-ik_{//}.r_{//}},
\end{array}
\eqno(10)
$$
where $k_{//}$ is the wave vector of electron in the x-y plane, $r_{//}=
(x,y)$.
Putting Eqs.(9)-(10) into Eqs.(1) and (3)-(4), we obtain the second
quantizating effective Hamiltonian
$$
H=\sum_{k_{//}}\f{\hbar^2{k_{//}}^2}{2m}{c^{\dag}}_{k_{//}}c_{k_{//}}
  +\f{p_z^2}{2m}+V(z)+
   \sum_q\hbar{\om}_{LO}{a_q}^{\dag}a_q
$$
$$
+\sum_{q_{//}}\hbar{\om}_{\pm}{a_{q_{//}}}^{\dag}a_{q_{//}}
+H_{e-BO}+H_{e-IO},
\eqno(11)
$$
where
$$
H_{e-BO}=\sum_{k_{//},q_{//}}\Gamma_{Ln}(q_{//},q_z,z)
{c^{\dag}}_{k_{//}+q_{//}}c_{k_{//}}
[{a^{\dag}}_{q_{//}}+a_{-q_{//}}],
\eqno(12)
$$
$$
H_{e-IO}=\sum_{k_{//},q_{//}}\Gamma_{IO}^{\pm}(q_{//},z)
{c^{\dag}}_{k_{//}+q_{//}}c_{k_{//}}
[{a^{\dag}}_{q_{//}}+a_{-q_{//}}].
\eqno(13)
$$
\hspace*{0.5cm}
In the following we shall derive the self-energy and effective mass
of an electron by using the
standard Matsubara Green function method[19].
For the weak-electron-phonon-coupling system, it is a good approximation that
we only take the first term in the perturbation series for the self-energy.
The contribution of
electron interacting with bulk optical phonons to the electron self-energy is
$$
\S_{Ln}({\bf{k}}_{//},ik_n)
=-\sum_{q_z>0}\sum_{q_{//}}
  \left[\Gamma_{Ln}(q_{//},q_z,z)\right]^2
  ({\bf{q}}_{//},q_j)\f{1}{\beta}
  \sum_{iq_n}{\cal{G}}^{(0)}(k_{//}+q_{//},ik_n+iq_n)
  {\cal{D}}^{(0)}(q_{//},iq_n)
$$
$$
=\sum_{q_z>0}\sum_{q_{//}}
  \f{2\pi \hbar\om_{Ln}e^2}{SL_z}
  \left[\f{1}{\eps_{\infty n}}-\f{1}{\eps_{\infty n}}\right]
  \f{\theta_n(z)sin^2(q_zz)}{{q_{//}}^2+{q_z}^2}
$$
$$
\times
\left(\f{n_p+n_F}{ik_n+\hbar{\omega}_{LO}-\var_{k_{//}+q_{//}}}+
\f{n_p+1-n_F}{ik_n-\hbar{\omega}_{LO}-\var_{k_{//}+q_{//}}}\right)
\eqno(14)
$$
with
$$
n_p=\f{1}{e^{\beta\hbar\omega_{LO}}-1},\hspace*{1cm}
n_F=\f{1}{e^{{\beta}\var_{k_{//}+q_{//}}}+1},
$$
where $n_p$ and $n_F$ are the phonon and fermion occupation factor,
${\cal{G}}^{(0)}(k_{//},ik_n)$ and ${\cal{D}}^{(0)}(q_{//},iq_n)$  are
the Green function of free electron and
phonon, respectively.
Set $ik_n=E+i\delta$ so that the real part of the retarded self-energy
of electron interacting with longitudinal optical(LO) phonons is
$$
Re[{\S_{Ln}}^{ret}(k_{//},E)]
=\f{\hbar\om_{Ln}e^2}{2}
\left(\f{1}{\eps_{\infty n}}-\f{1}{\eps_{\infty n}}\right)
\sum_{q_z>0}\f{1}{L_z}
{\int\limits_0}^{2\pi}{\int\limits_0}^{q_m}
\f{\theta_n(z)sin^2(q_zz)d^2q_{//}}{{q_{//}}^2+{q_z}^2}
$$
$$
 \times\left(\f{n_p+n_F}{E+\hbar\omega_{LO}-\var_{k_{//}+q_{//}}}
+\f{n_p+1-n_F}{E-\hbar\omega_{LO}-\var_{k_{//}+q_{//}}}\right).
$$
At zero temperature, $n_p=0$. The fermion occupation factors $n_F$ are all
zero, if there is only one particle in a band. After setting $k_{//}=0$, one can obtain
$$
Re[{\S_{Ln}}^{ret}(0,E)]=
\f{\hbar\om_{Ln}e^2}{8\pi}
\left(\f{1}{\eps_{\infty n}}-\f{1}{\eps_{\infty n}}\right)
$$
$$
{\int\limits_0}^{q_m}
\f{\theta_n(z)(1-cos(2q_zz))dq_z}{2(\hbar\om_{\pm}-E-E_{q_z})}
ln\f{(\hbar\om_{\pm}+x_m-E)E_{q_z}}{(\hbar\om_{\pm}-E)(x_m+E_{q_z})},
\eqno(15)
$$
where
$$
E_{q_z}=\f{{\hbar}^2{q_z}^2}{2m}.\hspace*{1cm}
\eqno(16)
$$
Considering the boundary of the
first Brillouin zone, the upper integral limit in above equations
can be written as
$$
q_m=\f{\sqrt{3}\pi}{a},\hspace*{1cm}
x_m=\f{{\hbar}^2{q_m}^2}{2m}=\f{3{\pi}^2{\hbar}^2}{2ma^2}.
\eqno(17)
$$
\hspace*{0.5cm}
The self-energy of an electron interacting with interface optical phonon is
$$
{\S_{l'}}^{IO}({\bf{k}}_{//},ik_n)
=-\sum_{q_{//}}
  \left[\Gamma_{IO}^n(q_{//},z)\right]^2
  ({\bf{q}}_{//},q_j)\f{1}{\beta}
  \sum_{iq_n}{\cal{G}}^{(0)}(k_{//}+q_{//},ik_n+iq_n)
  {\cal{D}}^{(0)}(q_{//},iq_n).
$$
When $k_{//}=0$, we get the real part of ${\S_{l'}}^{IO}({\bf k}_{//},ik_n)$
in the zero-temperature case
$$
{\S_{IO}}^{(1)}(0,E)
=\f{\hbar\om_{\pm}e^2}{4\pi}
\f{1}{{\beta_1}^{-1}(\om_{\pm})+{\beta_1}^{-1}(\om_{\pm})}
{\int\limits_0}^{q_m}\f{e^{-2q_{//}z}dq_{//}}
{E-\hbar{\omega}_{\pm}-\var_{q_{//}}},
\eqno(18)
$$
where
$$
\var_{q_{//}}=\f{\hbar^2{q_{//}}^2}{2m_1}.
\eqno(19)
$$
\hspace*{0.5cm}
The approximation result of effective mass can be got using
$$
\f{m}{m^*}=\lim_{k_{//}\rightarrow 0}\f{1+
             \f{\partial}{{\partial}\var_{k_{//}}}Re[\S^{ret}(k_{//},E)]}
             {1-\f{\partial}{{\partial}E}Re[\S^{ret}(k_{//},E)]}
\eqno(20)
$$
and
$$
          Re[\S^{ret}(k_{//},E)]=
              Re[{\S_{BO}}^{ret}(k_{//},E)]+Re[{\S_{IO}}^{ret}(k_{//},E)]
\eqno(21)
$$
{\normalsize\bf 3. Results and discussions}\\
\hspace*{0.5cm}
In this paper, we select the $GaAs/AlAs$ heterostructure as an example to study the polaron effect. The characteristic parameter concerned for $GaAs/AlAs$
heterostructure are taken as:
\begin{center}
Table 1: Material parameters of GaAs and AlAs\\
\begin{tabular}{lllllll}
\hline
materials & lattice constant(\AA) & $\eps_0$ & $\eps_{\infty}$ & $m_b/m_e$ &
$\hbar\om_{LO}(meV)$ & $\hbar\om_{TO}(meV)$\\
\hline
GaAs & 5.654 & 13.18 & 10.89 & 0.0656 & 36.25 & 33.29\\
AlAs & 5.654 & 10.06 & 8.16 & 0.147 & 50.09 & 44.88\\
\hline
\end{tabular}
\end{center}
in above table, $m_e$ is the mass of a free electron. \\
\hspace*{0.5cm}
In Figure 1 and 2, we take the electron self-energy and effective mass
in a single heterostructure as a function of distance of electron to
interface, respectively.
Here the parameters E and $k_{//}$ are set to 0.
Figure 1 shows that
the contribution of electron interacting with the half 3D bulk
 LO-phonon to the electron self-energy makes the electron self-energy
monotonically decrease starting from zero to that in bulk 3D
GaAs(AlAs) as the distance of electron to interface increases.
Figure 1 also show that
the contribution of electron interacting with interface-optical(IO)-phonon
to the electron self-energy makes the electron self-energy
monotonically increase starting from that(about -9.68meV)
in $GaAs/AlAs$ interface
to approaching to zero as the distance of electron to interface increases.
If we incorporate the contribution of electron interacting with the IO-
and half 3D bulk LO-phonon to the electron self-energy, we
find that the electron self-energy monotonically increases
starting from that(about -9.68meV) in $GaAs/AlAs$ interface to
that in 3D bulk GaAs(AlAs)
as the distance of electron to interface increases.\\
\hspace*{0.5cm}
Figure 2 shows that
the contribution of electron interacting with the half 3D bulk
 LO-phonon to the effective mass makes the effective mass
monotonically increase starting from zero to that in 3D bulk
GaAs(AlAs) as the distance of electron to interface increases.
Figure 2 also show that
the contribution of electron interacting with interface-optical(IO)-phonon
to the effective mass makes the effective mass
monotonically decrease starting from that($\Delta m^*/m_1$ is equal to about 0.04)
in $GaAs/AlAs$ interface
to approaching to zero with the increase of distance of electron to interface.
If we incorporate the contribution of electron interacting with the IO-
and half 3D bulk LO-phonon to the effective mass, we
find that the effective mass monotonically decreases
starting from that($\Delta m^*/m_1$ is equal to about 0.04)
in $GaAs/AlAs$ interface to that in 3D bulk GaAs(AlAs) crystal with the increase of the distance between the position of the electron and interface .
 When the distance of
an electron to interface is equal to 30 times of lattice constant, the
correction to effective mass is about 25 percent larger than that
in 3D bulk material.\\
{\normalsize\bf 4. Summary}\\
\hspace*{0.5cm}
We have studied the self-energy and effective mass of electron
in a single heterostructure using the Green's function method.
 In the theory, we expand the wave function in terms of
a basis set which is only on the plane paralleling to interface 
so as to put Hamiltonian into the second quantization representation only on
plane paralleling to interface.
 Numerical calculations using $GaAs/AlAs$ heterostructure as
an example is performed.
The results show that the total self-energy,
which incorporate the contribution of electron interacting with
interface-optical phonons and bulk LO
phonons to electron self-energy, 
monotonically increase starting from
that in  $GaAsAl/As$ interface to that in 3D bulk $GaAs(AlAs)$ as the
 distance of electron to interface increase.
The results also show that the band mass,
which incorporate the contribution of electron interacting with
interface-optical phonons and 3D bulk LO
phonons to the band mass, 
monotonically decrease starting from
that in  $GaAs/AlAs$ heterostructure to that in 3D bulk $GaAs(AlAs)$ as the
 distance of electron to interface increase. One point the authors want to make clear 
is that in the present paper we aims at studying the polaron effect on the electron mass and self-energy. The potential in the z direction consist of V(z) and the electron-phonon interaction 
induced parts $\S_{BO}(k_{//},E)$ and  $\S_{IO}(k_{//},E)$ . The z-direction electron wave function is rather a complicated problem, which needs further investigation . One can also
see that the one phonon process electron self-energy include the recoil part of the electron 
,which is the advantage of Green function method upon the second perturbation theory (for instance 
compared with Ref [17]). \\
{\bf Acknowledgment}\\
\hspace*{0.5cm} The present work was supported by the National Nature Science Foundation of China under Grant No.10174049. 

\newpage

\begin{center}
        {\bf  Figures Captions}
\end{center}
Figure 1. The self-energy of the polaron
 in $GaAs/AlAs$ heterostructure
 as a function of distance between the position of the electron and the interface
 for $E$ and $k_{//}=0$.
\vspace*{1cm}\\
{\bf Figure 2.} The effective mass of the polaron
 in $GaAs/AlAs$ heterostructure
 as a function of distance  between the position of the electron and the interface
 for $E$ and $k_{//}=0$. $m_1$ and $m_2$ 
are the electron band masses in the GaAs and AlAs bulk materials, respectively.

\newpage
\vfil\includegraphics[scale=0.7]{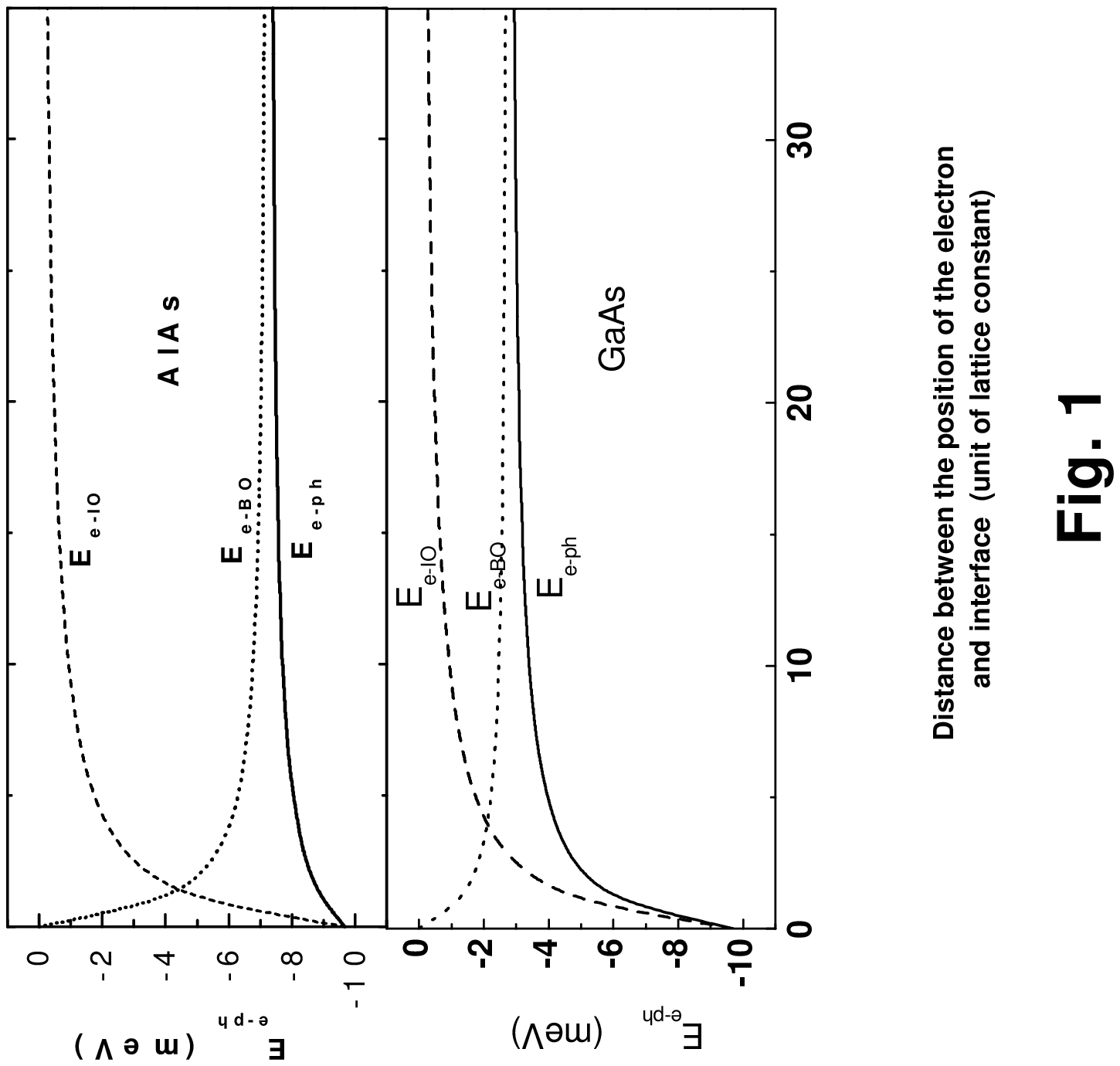}\vfil

\newpage
\vfil\includegraphics[scale=0.7]{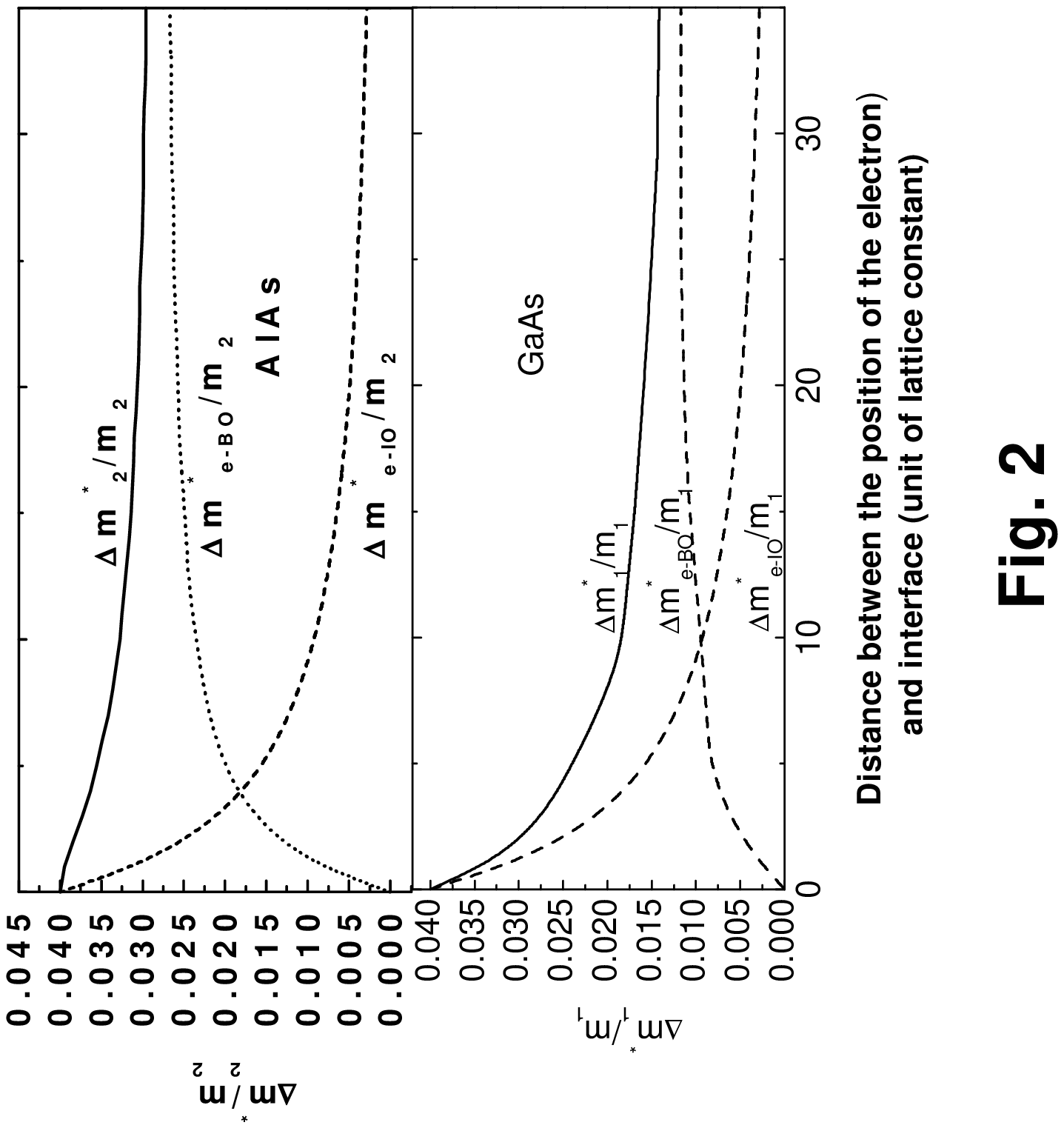}\vfil

\end{document}